\documentclass{desyproc}

\begin{document}
\newcommand{\GPDtE}{\langle \tilde{E} \rangle}
\newcommand{\GPDtH}{\langle \tilde{H} \rangle}
\newcommand{\GPDHT}{\langle H_T \rangle}
\newcommand{\GPDETbar}{\langle \bar{E}_T \rangle}

\title{Deeply pseudoscalar meson electroproduction with CLAS and Generalized Parton Distributions}

\author{{\slshape Michel Guidal$^1$, Valery Kubarovsky$^2$}\\[1ex]
$^1$Institut de Physique Nucl\'{e}aire Orsay, CNRS-IN2P3, Universit\'e Paris-Sud, France\\
$^2$Thomas Jefferson National Accelerator Facility, Newport News, Virginia 23606}

\contribID{xy}

\confID{8648}  
\desyproc{DESY-PROC-2014-04}
\acronym{PANIC14} 
\doi  

\maketitle

\begin{abstract}
We discuss the recent data of exclusive $\pi^0$ (and $\pi^+$) electroproduction on the proton
obtained by the CLAS collaboration at Jefferson Lab. It is observed that the 
cross sections, which have been decomposed in
$\sigma_T+\epsilon\sigma_L$, $\sigma_{TT}$ and $\sigma_{LT}$ structure functions,
are dominated by transverse amplitude contributions. The data can be interpreted in the
Generalized Parton Distribution formalism provided that one includes
helicity-flip transversity GPDs.
\end{abstract}

\begin{wrapfigure}{r}{0.45\textwidth} 
\vspace*{-30pt}
\begin{center}
\includegraphics[width=0.4\textwidth]{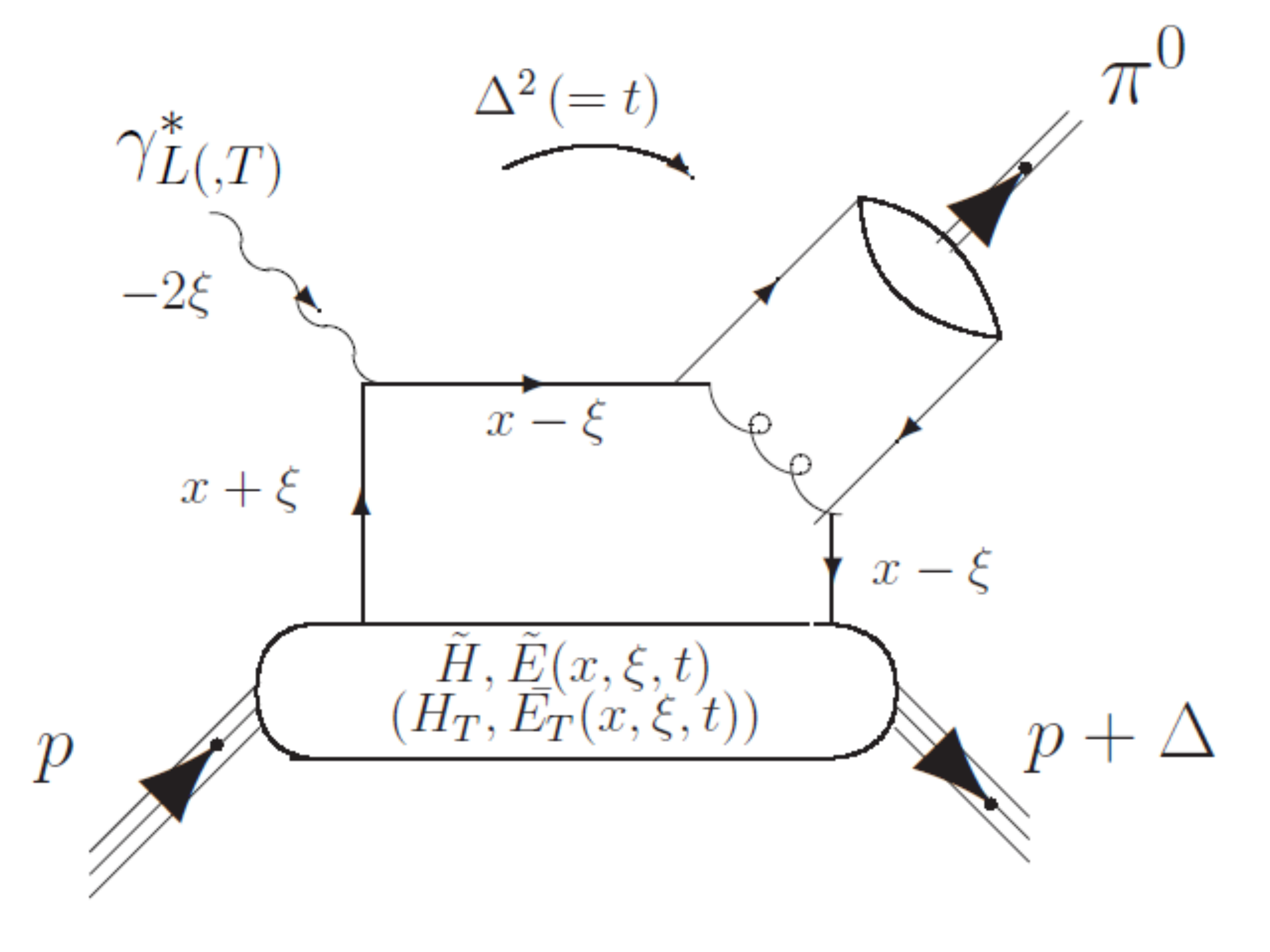}
\caption{The ``handbag" diagram for exclusive $\pi^0$ electroproduction on the proton
in terms of GPDs. When longitudinal photons are involved, only the helicity-conserving
GPDs $\tilde{H}$ and $\tilde{E}$ enter, while for transverse photons, the 
helicity-flip GPDs $H_T$ and $\bar{E_T}$ also enter the process. The various 
kinematical variables are explained in the text.}
\label{fig:diagps}
\end{center}
\vspace*{-15pt}
\end{wrapfigure} 

The formalism of Generalized Parton Distributions (GPDs) which has appeared 
in the last two decades (Refs.~\cite{Mueller:1998fv,Ji:1996ek,Radyushkin:1996nd}
for the original articles and Ref.~\cite{Guidal:2013rya} for a recent review)
allows to interpret the exclusive electroproduction of photons or mesons on the nucleon
in terms of quarks and gluons (i.e. partons), the fundamental degrees of freedom of 
Quantum Chromodynamics (QCD).
It has been shown~\cite{Collins:1996fb} that for these processes, at sufficiently large virtuality
of the photon $Q^2=(e-e^\prime)^2$, there is a factorization between a ``hard" 
elementary scattering part at the quark or gluon level, 
exactly calculable in perturbative QCD, and a non-perturbative nucleon 
structure part, which encodes all the complex partonic structure of the nucleon
and which is parametrized by GPDs.
This factorization is illustrated in Fig.~\ref{fig:diagps} for the case of $\pi^0$
electroproduction on the proton, on which we will focus in this article. For pseudoscalar meson production, it 
is shown that, at leading-twist QCD, this factorization is valid only 
for longitudinal
incoming photons, that the longitudinal part of the cross section $\sigma_L$ should
dominate at asymptotically large $Q^2$ valuse and that two quark helicity-conserving GPDs contribute to the 
process: $\tilde H$ and $\tilde E$.
These two GPDs correspond to the amplitudes where the nucleon spin remains unchanged or has been
flipped respectively. At QCD leading-order, the GPDs depend on three independent variables: $x$, $\xi$ and
$t$. In simple terms, GPDs represent, in a frame where the nucleon goes to the speed of light 
in a certain direction, the probability amplitude of finding a quark in the
nucleon with a longitudinal momentum fraction $x +\xi$ and of putting it back into the
nucleon with a different longitudinal momentum fraction $x -\xi$, plus some transverse
momentum ``kick", which is represented by $t$. For the particular case
of $\xi=0$, the momentum transfer $\Delta$ (with $\Delta^2=t$) is the conjugate variable of the impact parameter 
$b_\perp$ so that the GPDs encode both the longitudinal momentum distributions of
partons inside the nucleon through their dependence on $x$ and their transverse position distributions 
through their dependence on $t$. This allows for a sort of tomography of the nucleon
where one can probe the transverse size of the nucleon for different quark momentum slices.

\begin{figure*}[ht]
\begin{center}
\includegraphics[width=0.75\textwidth]{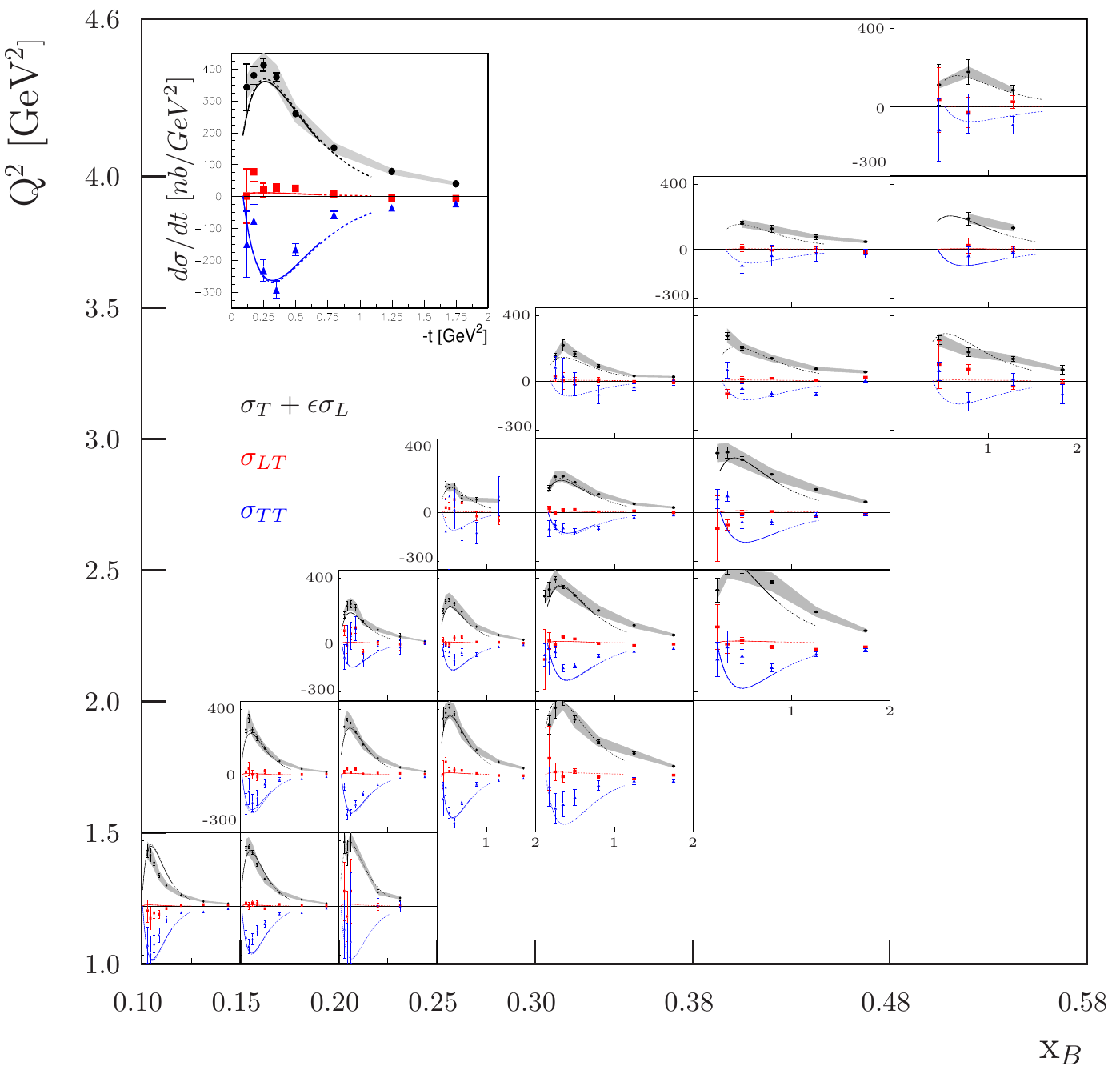}
\end{center}
\caption{ \label{fig:GK-GL} 
The extracted structure functions vs. $t$ as measured by CLAS. The data and curves are as follows: 
 black (filled circles)  - $d\sigma_U/dt =d\sigma_T/dt +\epsilon d\sigma_L/dt$,\  blue (triangles) - 
 $d\sigma_{TT}/dt$ ,  and  red (squares) - $d\sigma_{LT}/dt$.
 The curves are theoretical predictions produced  with the models of Refs.~\cite{G-K-11} (solid) and 
 ~\cite{Goldstein:2010gu} (dashed).
}
\end{figure*}

Recently, the CLAS collaboration has measured at Jefferson Lab with a 5.75-GeV electron beam
the 4-fold differential cross sections $d^4\sigma/dtdQ^2dx_Bd\phi_\pi$ 
\footnote{The standard $x_B$ Bjorken variable is related to $\xi$: $\xi\simeq x_B/(2-x_B)$ and $\phi_\pi$ is the azimuthal angle between 
the leptonic and hadronic planes.} of the
$ep\to ep\pi^0$ reaction, thus extracting the structure 
functions $\sigma_T+\epsilon\sigma_L, \sigma_{TT}$ and $\sigma_{LT}$ as functions of $t$
over a wide range of $Q^2$ and $x_B$~\cite{Bedlinskiy:2014tvi}. Fig.~\ref{fig:GK-GL}
shows a sample of these results (1800 kinematic points in bins of $Q^2, x_B$, 
$t$ and $\phi_\pi$ were measured in all). These results are 
in agreement with the results of Ref.~\cite{Collaboration:2010kna}, 
which published  high  accuracy cross sections in a limited kinematical range
in the lower $Q^2$, $W$ and $|t|$ regions of the present experiment. One observes that the $d\sigma_{TT}/dt$ structure 
function (which is negative)
is comparable in magnitude with the unpolarized structure function $d\sigma_U/dt=
d\sigma_T/dt+\epsilon d\sigma_L/dt$).
Furthermore, $d\sigma_{LT}/dt$ is small in comparison with $d\sigma_U/dt$ and $d\sigma_{TT}/dt$.
In the same vein, in an earlier CLAS measurement~\cite{demasi},
sizeable beam-spin asymmetries (proportional to the fifth structure function $\sigma_{LT^\prime}$),
were found for this same channel. Such non-zero asymmetries 
imply that both transverse and longitudinal amplitudes participate in the process.
Similarly, at higher energies, the HERMES collaboration measured the transverse target spin asymmetry 
in the ``cousin" channel of $\pi^+$ electroproduction~\cite{hermes-transverse}. The sizeable results can
also only be explained by significant transverse amplitude contributions.

All these experimental observations point to the model-independent conclusion that the asymptotic 
leading-order handbag approach 
for which the longitudinal part of the cross section is dominant is not applicable at the present 
values of $Q^2$. Although model-dependent, this is confirmed by theoretical calculations 
of the handbag diagram for longitudinal virtual photons based solely on the $\tilde H$ and $\tilde E$ GPDs 
which are found to underestimate the measured cross sections by more than an order of magnitude, even after 
including  finite--size  corrections through Sudakov form factors\cite{G-K-09}.

This failure to describe these experimental results for exclusive pseudo-scalar meson electroproduction
with quark helicity-conserving GPDs recently stimulated the consideration of the role of the 
chiral-odd quark helicity-flip contributions (i.e. where the active quark in Fig.~\ref{fig:diagps}
undergoes a helicity-flip), in particular through the introduction of so-called
transversity GPDs; namely: $H_T$, which characterizes the quark distributions involved in 
nucleon helicity-flip, and $\bar E_T (= 2\widetilde H_T + E_T) $ which characterizes the quark distributions 
involved in  nucleon  helicity-non-flip processes~\cite{diehl_haegler, Goekeler}.

Pseudoscalar meson electroproduction, and in particular $\pi^0$ production, 
was identified~\cite{G-K-09,Ahmad:2008hp,G-K-11} as especially 
sensitive to the quark helicity-flip subprocesses. The produced meson has no intrinsic helicity 
so that the angular momentum of the incident photon is either transferred to the nucleon via a quark 
helicity-flip or involves  orbital angular momentum processes. In addition, for $\pi^0$ 
production the structure of the amplitudes further suppresses the quark helicity-conserving 
amplitudes relative to the  helicity-flip amplitudes~\cite{G-K-09}. 

The results of two GPD-based models which include transversity GPDs \cite{G-K-11,Goldstein:2010gu}  are  
superimposed in Fig.~\ref{fig:GK-GL}. The GL and GK approaches, though employing different models of 
GPDs, lead to transverse photon amplitudes that are much larger than the longitudinal amplitudes.
These latter account for only a small fraction (typically less than 10\% ) of the unseparated  
structure functions $d\sigma_T/dt+ \epsilon d\sigma_L/dt$ in the kinematic regime under investigation. 
With such inclusion of the quark-helicity non-conserving  chiral-odd GPDs, which  contribute primarily to 
$d\sigma_T/dt$ and  $d\sigma_{TT}/dt$  and, to a lesser extent, to $d\sigma_{LT}/dt$, the model of 
Ref.~\cite{G-K-11} agrees rather well with the data. Deviations in shape become greater at smaller 
$-t$ for the unseparated cross section $d\sigma_U/dt$. The behavior of the cross section as 
$|t| \to |t|_{min}$ is determined by the interplay between $H_T$ and $\bar E_T$. 
For the GPDs of Ref.~\cite{G-K-11}  the parameterization was guided by the lattice calculation 
results of Ref.~\cite{Goekeler},  while Ref.~\cite{Goldstein:2010gu} used a GPD Reggeized 
diquark-quark model to obtain the GPDs. The results  in Fig.~\ref{fig:GK-GL} 
for  the model of Ref.~\cite{G-K-11} (solid curves), in which  $\bar E_T$  is dominant,  agree rather well with 
the data. In particular, the  structure function $\sigma_U$ begins to decrease as $|t |\to |t|_{min}$, showing 
the effect of $\bar E_T$.    In the model of 
Ref.~\cite{Goldstein:2010gu}  (dashed curves) $H_T$ is dominant, which leads to a large rise in cross section as
$-t$ becomes small so that the contribution of $\bar E_T$  relative to $H_T$ appears to be underestimated.
One can make a similar  conclusion from  the comparison between data and model predictions for  
$\sigma_{TT}$. This  shows the sensitivity of the measured  $\pi^0$ structure functions  for constraining  
the transversity GPDs.

We also mention that $\pi^+$ electroproduction has also been measured 
by the CLAS collaboration~\cite{Park:2012yf} in the same phase space.
It is found that the GK model describes also qualitatively the low-$t$ 
unseparated cross sections over the whole ($x_B$, $Q^2$) domain, when the same
transversity GPDs are included. In $\pi^+$ production, the role of transversity GPDs
is less apparent because of the presence and dominance of the longitudinal $\pi^+$-pole 
term (which is absent in $\pi^0$ production). However, this latter contribution has an important contribution 
only in the low $|t|$ domain and only for the lowest $x_B$ and the largest $Q^2$ values,
leaving sensitivity to other contributions, namely transversity GPDs.
 
In conclusion, differential cross sections of exclusive $\pi^0$ (and $\pi^+$) electroproduction 
on the proton have been obtained in the few-GeV region in a wide $Q^2, x_B, t, \phi_\pi$ phase space 
with the CLAS detector at JLab, from which the structure functions  $d\sigma_U/dt$, 
$d\sigma_{TT}/dt$ and $d\sigma_{LT}/dt$ could be extracted. 
It is found that $d\sigma_U/dt$ and $d\sigma_{TT}/dt$ are comparable in magnitude with each other, while 
$d\sigma_{LT}/dt$ is very much smaller than either pointing to the dominance
of transverse amplitude contributions to the process.

Within the handbag interpretation, there are two independent theoretical calculations~\cite{G-K-11, 
Goldstein:2010gu} which confirm that the  measured unseparated cross sections 
are much larger than expected from leading-twist handbag calculations which 
are dominated by longitudinal photons. When including transversity GPDs, the general shapes
and magnitudes of the various structure functions are reproduced. Extensive new CLAS measurements 
of beam spin, target spin and double-spin asymmetries for exclusive
pseudo-scalar electroproduction on the proton are currently under analysis. Comparison of 
these results  with theoretical models will allow to confirm (or not) the GPDs interpretations that we
have outlined here.

This material is based upon work supported by the U.S. Department of Energy,
Office of Science, Office of Nuclear Physics under contract
DE-AC05-06OR23177.

\begin{footnotesize}

\end{footnotesize}



\begin{thebibliography}{99}
\bibitem{Mueller:1998fv} 
  D. Mueller, D. Robaschik, B. Geyer, F. M. Dittes and J. Horejsi,  
  Fortsch.\ Phys.  {\bf 42} (1994) 101. 
  
\bibitem{Ji:1996ek} 
  X. D. Ji,
  Phys.\ Rev.\ Lett.  {\bf 78} (1997) 610, Phys.\ Rev.\ D {\bf 55} (1997) 7114.


\bibitem{Radyushkin:1996nd}
  A. V. Radyushkin,
  Phys.\ Lett.\  B {\bf 380} (1996) 417, Phys.\ Rev.\ D {\bf 56} (1997) 5524. 




\bibitem{Guidal:2013rya} 
  M.~Guidal, H.~Moutarde and M.~Vanderhaeghen,
  Rept.\ Prog.\ Phys.\  {\bf 76} (2013) 066202.

\bibitem{Collins:1996fb}
  J.~C.~Collins, L.~Frankfurt and M.~Strikman,
  Phys.\ Rev.\ D {\bf 56} (1997) 2982.
  
  \bibitem{Bedlinskiy:2014tvi}
  I.~Bedlinskiy {\it et al.}  [CLAS Collaboration],
  arXiv:1405.0988 [nucl-ex].

\bibitem{Collaboration:2010kna}
  E.~Fuchey, A.~Camsonne, C.~Munoz Camacho, M.~Mazouz, G.~Gavalian, E.~Kuchina, M.~Amarian and K.~A.~Aniol {\it et al.},
  Phys.\ Rev.\ C {\bf 83} (2011) 025201.

\bibitem{demasi} 
  R.~De Masi {\it et al.}  [CLAS Collaboration],
  Phys.\ Rev.\ C {\bf 77} (2008) 042201.

\bibitem{hermes-transverse} 
  A.~Airapetian {\it et al.}  [HERMES Collaboration],
  Phys.\ Lett.\ B {\bf 682} (2010) 351.

\bibitem{G-K-09}
  S.~V.~Goloskokov and P.~Kroll,
  Eur.\ Phys.\ J.\ C {\bf 65} (2010) 137.

\bibitem{diehl_haegler}
  M.~Diehl and P.~Hagler,
  Eur.\ Phys.\ J.\ C {\bf 44} (2005) 87.

\bibitem{Goekeler}
  M.~Gockeler {\it et al.}  [QCDSF and UKQCD Collaborations],
  Phys.\ Rev.\ Lett.\  {\bf 98} (2007) 222001.
  
  \bibitem{Ahmad:2008hp}
  S.~Ahmad, G.~R.~Goldstein and S.~Liuti,
  Phys.\ Rev.\ D {\bf 79} (2009) 054014.

\bibitem{G-K-11}
  S.~V.~Goloskokov and P.~Kroll,
  Eur.\ Phys.\ J.\ A {\bf 47} (2011) 112.

\bibitem{Goldstein:2010gu}  
  G.~R.~Goldstein, J.~O.~Hernandez and S.~Liuti,
  Phys.\ Rev.\ D {\bf 84} (2011) 034007,
  Int.\ J.\ Mod.\ Phys.\ Conf.\ Ser.\  {\bf 20} (2012) 222,
  J.\ Phys.\ G {\bf 39} (2012) 115001.

\bibitem{Park:2012yf}
  K.~Park {\it et al.}  [CLAS Collaboration],
  Phys.\ Rev.\ C {\bf 85} (2012) 035208.

\end{thebibliography}
\end{document}